\documentclass[conference]{IEEEtran}
\IEEEoverridecommandlockouts
\usepackage{cite}
\usepackage{amsmath,amssymb,amsfonts}
\usepackage{algorithmic}
\usepackage{graphicx}
\usepackage{textcomp}
\usepackage{xcolor}
\usepackage{multirow}
\usepackage[super]{nth}
\def\BibTeX{{\rm B\kern-.05em{\sc i\kern-.025em b}\kern-.08em
    T\kern-.1667em\lower.7ex\hbox{E}\kern-.125emX}}
\usepackage[ruled,vlined, linesnumbered]{algorithm2e}

\begin{document}

\title{Decoding Fatigue Levels of Pilots Using EEG Signals with Hybrid Deep Neural Networks
\footnote{{\thanks{This research was supported by the Challengeable Future Defense Technology Research and Development Program through the Agency For Defense Development (ADD) funded by the Defense Acquisition Program Administration (DAPA) in 2024 (No.912911601) was partly supported by the Institute of Information \& Communications Technology Planning \& Evaluation (IITP) grant, funded by the Korea government (MSIT) (No. RS-2019-II190079, Artificial Intelligence Graduate School Program (Korea University)).}
}}
}

\author{
\IEEEauthorblockN{Dae-Hyeok Lee}
\IEEEauthorblockA{\textit{Dept. of Brain and Cognitive Engineering} \\
\textit{Korea University} \\ 
Seoul, Republic of Korea \\
lee\_dh@korea.ac.kr}

\and

\IEEEauthorblockN{Sung-Jin Kim}
\IEEEauthorblockA{\textit{Dept. of Artificial Intelligence} \\
\textit{Korea University} \\
Seoul, Republic of Korea \\
s\_j\_kim@korea.ac.kr}

\and

\IEEEauthorblockN{Si-Hyun Kim}
\IEEEauthorblockA{\textit{Dept. of Artificial Intelligence} \\
\textit{Korea University} \\
Seoul, Republic of Korea \\
kim\_sh@korea.ac.kr}
}

\maketitle

\begin{abstract}
The detection of pilots' mental states is critical, as abnormal mental states have the potential to cause catastrophic accidents. This study demonstrates the feasibility of using deep learning techniques to classify different fatigue levels, specifically a normal state, low fatigue, and high fatigue. To the best of our knowledge, this is the first study to classify fatigue levels in pilots. Our approach employs the hybrid deep neural network comprising five convolutional blocks and one long short--term memory block to extract the significant features from electroencephalography signals. Ten pilots participated in the experiment, which was conducted in a simulated flight environment. Compared to four conventional models, our proposed model achieved a superior grand--average accuracy of 0.8801 ($\pm$0.0278), outperforming other models by at least 0.0599 in classifying fatigue levels. In addition to successfully classifying fatigue levels, our model provided valuable feedback to subjects. Therefore, we anticipate that our study will make the significant contributions to the advancement of autonomous flight and driving technologies, leveraging artificial intelligence in the future.
\end{abstract}


\section{INTRODUCTION}
Brain--computer interface (BCI) enables interaction between humans and devices by interpreting user's cognitive status and intentions \cite{kim2024towards, mane2020multi, abibullaev2021systematic}. BCIs are generally categorized into invasive and non--invasive types \cite{kim2015abstract}. Invasive BCIs record neural activity through implanted electrodes positioned near target neurons \cite{ahn2022multiscale}. This approach offers the benefit of a high signal--to--noise ratio (SNR) due to direct brain activity measurement but necessitates surgical intervention. Conversely, non--invasive BCIs acquire brain activity data without the need for surgery \cite{liu2021assessing, prabhakar2020framework}. While this method avoids surgical procedures and is more cost--effective, it suffers from a relatively lower SNR \cite{ahn2022multiscale}. As a result, non--invasive BCIs have been utilized in various applications, including the control of external devices such as drones \cite{lee2021design}, robotic arms \cite{meng2016noninvasive}, wheelchairs \cite{kim2018commanding}, and spellers \cite{yin2014dynamically}.

In the field of BCI, the accurate detection of humans' abnormal mental states with high performance remains a critical concern \cite{lee2023autonomous}. Recently, numerous technologies related to autonomous flight and driving have been developed for application in real--world environments. Given that pilots' or drivers' mental state is a key factor in ensuring passenger safety, the ability to reliably detect these states is crucial for enhancing safety measures. Flight operation is a demanding task due to the significant energy expenditure involved \cite{yen2009investigation}. Fatigue is commonly induced by extended cognitive tasks, particularly those that are repetitive or monotonous \cite{lee2023autonomous}. According to the British Airline Pilots' Association, 280 out of 500 commercial pilots reported having fallen asleep during night flights.

In the context of detecting humans' mental states, electroencephalography (EEG) signals are highly informative as they directly reflect an individual's cognitive status and intentions \cite{jeong2019classification}. Several studies have focused on detecting fatigue using solely EEG signals. Wu \textit{et al}. \cite{wu2021fatigue} proposed the nonparametric prior--induced deep sum--logarithmic--multinomial mixture model for identifying pilots' fatigue through the development of a brain power map, along with the adaptive topic--layer stochastic gradient Riemann Markov chain Monte Carlo inference method for estimating global parameters without relying on heuristic assumptions. Yang \textit{et al}. \cite{yang2019complex} introduced the novel complex network--based broad learning system for the EEG--based fatigue detection. Their results demonstrated that the method could reliably distinguish between fatigue and alert states with high stability.

The contributions of this study are as follows: \textit{i}) We developed an experimental paradigm to acquire EEG signals related to fatigue in a flight environment involving pilots. We effectively induced a normal state (NS), low fatigue (LF), and high fatigue (HF) based on task variations. \textit{ii}) We introduced the novel model comprising five convolutional blocks and one long short--term memory (LSTM) block for classifying fatigue levels. Our proposed model demonstrated a superior accuracy compared to the conventional models. To the best of our knowledge, our study represents the first attempt to classify fatigue levels using the deep learning architecture.\\

\begin{figure}[t!]
\centering
\scriptsize
\centerline{\includegraphics[width=0.8\columnwidth, height=0.26\textwidth]{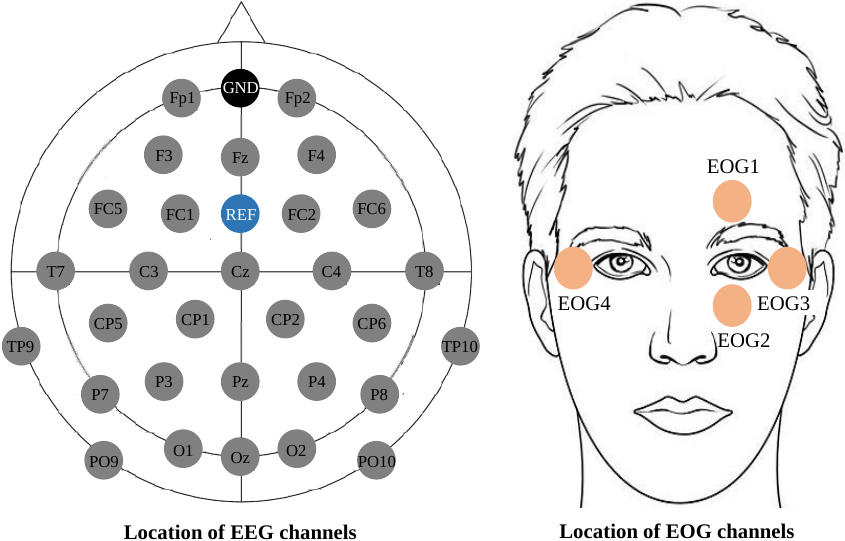}}
\caption{Data configuration of EEG and EOG channels.}
\end{figure}

\section{MATERIALS AND METHODS}
\subsection{Subjects}
We acquired EEG data related to fatigue from ten pilots (S1--S10, aged 25.6 ($\pm$0.52)). The recruitment criterion required subjects to have over 100 hr. of flight experience at the Taean Flight Education Center. All subjects had no history of neurological or psychiatric disorders. Prior to the experiment, the study protocol was thoroughly explained to subjects to ensure their understanding, and written informed consent was obtained in accordance with the Declaration of Helsinki. Subjects were also asked to complete a questionnaire to assess their mental and physical conditions before and after the experiment to evaluate the experimental paradigm. The experimental protocols were approved by the Institutional Review Board of Korea University [1040548--KU--IRB--18--92--A--2].\\

\subsection{Experimental Environment}
We utilized the Cessna 172 (Garmin, Olathe, KS) for the flight simulation, which included a screen, the cockpit, and signal amplifier. The cockpit was equipped with a flight yoke and various control panels to create a realistic flight environment. EEG and EOG signals were recorded using a signal amplifier (BrainAmp, Brain Products GmbH, Germany). The sampling rate for EEG and EOG signals was set at 1,000 Hz, with the 60 Hz notch filter applied to eliminate direct current noise. As shown in Fig. 1, 30 EEG channels were positioned to the subjects’ scalps according to the international 10/20 system, and four EOG channels were placed along the vertical and horizontal axes around the eyes. The AFz and FCz channels served as the ground and reference electrodes, respectively. Prior to the experiment, the impedance of all electrodes was reduced to below 10 k$\Omega$ by applying conductive gel to the subjects' scalps.\\

\subsection{Experimental Protocol and Paradigm}
We designed an experimental paradigm specifically to effectively induce fatigue in pilots, as shown in Fig. 2. The paradigm was meticulously designed, and the experiment was conducted under controlled conditions to ensure accurate acquisition of EEG data related to fatigue. Subjects were not given additional time to practice the tasks, as familiarity with the tasks could prevent the induction of intended abnormal mental state.

Subjects performed a monotonous nighttime flight simulation, with the initial settings configured to 3,000 feet (height), $0\,^{\circ}$ (heading), and 100 knots (velocity) for one hour to induce fatigue. A beep sound appeared every minute to indicate when subjects had to press a number on the keypad. The beep sound presented to subjects was soft, at 40 dB, rather than sharp or loud. Sounds below 48 dB are not known to induce wakefulness \cite{maschke2004nocturnal}, ensuring that the 40 dB sound did not interfere with the fatigue induction process. Subjects also provided input using the Karolinska sleepiness scale (KSS), which is a widely recognized index for assessing subjective fatigue. The KSS comprises 9 levels, with the specific score varying for each subject based on their subjective assessment of their fatigue levels.\\

\noindent
-- level 1: extremely alert\\
-- level 2: very alert\\
-- level 3: alert\\
-- level 4: rather alert\\
-- level 5: neither alert nor sleepy\\
-- level 6: some signs of sleepiness\\
-- level 7: sleepy, but no difficulty remaining alert\\
-- level 8: sleepy, some effort to keep alert\\
-- level 9: extremely sleepy\\

If subjects failed to input the KSS or were unable to operate the aircraft successfully, that period was classified as level 9. We categorized levels 1 to 3 as a normal state (NS), levels 4 to 6 as low fatigue (LF), and levels 7 to 9 as high fatigue (HF). Following the fatigue experiment, a questionnaire was administered to assess subjects' conditions, and all subjects reported feeling fatigued without experiencing any disturbances.\\

\begin{figure}[t!]
\centering
\scriptsize
\centerline{\includegraphics[width=\columnwidth, height=0.16\textwidth]{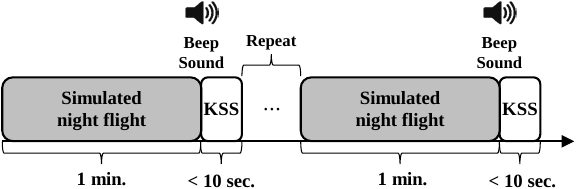}}
\caption{Experimental paradigm for inducing fatigue in a simulated flight environment.}
\end{figure}

\subsection{Signal Preprocessing}
Preprocessing of EEG signals was performed using the BBCI toolbox \cite{blankertz2010berlin} in MATLAB 2019a. The signals were band--pass filtered between 1 and 50 Hz using the $2^{nd}$ order zero--phase Butterworth filter and subsequently downsampled from 1,000 to 100 Hz. To enhance the quality of EEG data, the independent component analysis was employed to eliminate artifacts caused by unnecessary pilot movements, such as eye blinks and head movements. The EOG channels were used to identify and remove eye--related artifacts via ICA. The acquired data were segmented into 1 sec. epochs without overlap \cite{jeong2019classification}. As a result, 3,600 samples (60 samples$\times$60 trials) were obtained for each subject, yielding a total of 36,000 samples across all subjects.\\

\begin{table}[t!]
\centering
\caption{Comparison of an accuracy for classifying fatigue levels with the statistical analysis among the conventional models and the proposed model.}
\scriptsize
\renewcommand{\arraystretch}{2.0}
\resizebox{\columnwidth}{!}{
\begin{tabular}{cccccc}
\hline
Subject & PSD--SVM \cite{zhang2017design}        & DeepConvNet \cite{schirrmeister2017deep}    & EEGNet \cite{lawhern2018eegnet}         & MFB--CNN \cite{lee2020continuous}        & Proposed \\ \hline
S1      & 0.7008  & 0.7312      & 0.7618 & 0.7826  & 0.8519   \\
S2      & 0.7033  & 0.7403      & 0.7774 & 0.7926  & 0.8841   \\
S3      & 0.6315  & 0.7716      & 0.8003 & 0.8153  & 0.8796   \\
S4      & 0.7157  & 0.7698      & 0.7997 & 0.8139  & 0.8706   \\
S5      & 0.7558  & 0.8045      & 0.8506 & 0.8818  & 0.8799   \\
S6      & 0.7344  & 0.8191      & 0.8529 & 0.8783  & 0.9404   \\
S7      & 0.6069  & 0.7206      & 0.7601 & 0.7841  & 0.8626   \\
S8      & 0.7408  & 0.7988      & 0.8126 & 0.8316  & 0.8912   \\
S9      & 0.7256  & 0.7792      & 0.7991 & 0.8211  & 0.9003   \\
S10     & 0.6819  & 0.7534      & 0.7879 & 0.8006  & 0.8406   \\ \hline
Avg.    & 0.6997  & 0.7689      & 0.8002 & 0.8202  & 0.8801   \\
Std.    & 0.0478  & 0.0327      & 0.0320 & 0.0353  & 0.0278   \\ \hline
\textit{p}--value     & \textless{}0.05 & \textless{}0.05 & \textless{}0.05 & \textless{}0.05 & --        \\ \hline
\end{tabular}}
\end{table}

\subsection{The Proposed Model}
We proposed the deep learning--based model for classifying fatigue levels from EEG signals with high accuracy and robustness. EEG signals contain various features, including spectral, spatial, and temporal information. To effectively capture these features, we employed the hybrid deep learning framework. Our proposed model incorporates five convolutional blocks followed by one LSTM block.

The spatio--temporal convolutional neural network (CNN) in our proposed model consists of five convolutional blocks designed to extract spatial and spectral features from EEG signals. The $1^{st}$, the $2^{nd}$, and the $3^{rd}$ convolutional blocks each comprise two convolutional layers with 1$\times$5 filters, a 1$\times$1 stride, and a batch normalization layer, with batch sizes of 32, 64, and 128, respectively. The 1$\times$5 filter is used to capture temporal information. The $4^{th}$ and the $5^{th}$ convolutional blocks each contain three convolutional layers, with 5$\times$1 and 3$\times$1 filters, respectively, a 1$\times$1 stride, and a batch normalization layer, with 128 and 256 feature maps, respectively. The 5$\times$1 and 3$\times$1 filters are designed to extract spatial features. To mitigate overfitting, we applied both max--pooling and average--pooling layers. Additionally, in the $5^{th}$ convolutional block, the exponential linear unit (ELU) was used as the activation function.

The data processed by the $5^{th}$ convolutional block is subsequently fed into the LSTM block. The LSTM network, a type of recurrent neural network, is widely recognized for its effectiveness in detecting mental states. In our proposed model, the LSTM block consists of two LSTM layers, with 256 and 128 hidden units, respectively. This LSTM block is employed to extract essential temporal features from EEG signals.

The classification block represents the final component of our proposed model, comprising three fully connected layers and a softmax layer. The $1^{st}$ and the $2^{nd}$ fully connected layers contain 128 and 64 hidden units, respectively. The output of the $3^{rd}$ fully connected layer is passed to a 3--way softmax layer, which generates a class label distribution across three classes (NS, LF, and HF).\\

\section{RESULTS AND DISCUSSION}
\subsection{Performance Evaluation}
The five--fold cross--validation method was employed to ensure a fair evaluation of accuracy. The dataset was randomly shuffled and divided into five parts, with four parts used for training and one for validation. This process was repeated four times, each with a different shuffle order. We compared our model with several baseline models: the power spectral density--support vector machine (PSD--SVM) \cite{zhang2017design}, the DeepConvNet \cite{schirrmeister2017deep}, the EEGNet \cite{lawhern2018eegnet}, and the multiple feature block--based CNN (MFB--CNN) \cite{lee2020continuous}. The PSD–SVM \cite{zhang2017design} utilizes the power spectral density of the \textit{$\delta$}-- (1--4 Hz), \textit{$\theta$}-- (4--8 Hz), \textit{$\alpha$}-- (8--13 Hz), and \textit{$\beta$}--bands (13--30 Hz) as features, with the SVM serving as the classifier. The DeepConvNet \cite{schirrmeister2017deep} comprises four convolutional blocks, each including a convolutional layer, a batch normalization, and an ELU activation. The $1^{st}$ block extracts both spatial and temporal features, while the subsequent blocks focus solely on temporal features. The EEGNet \cite{lawhern2018eegnet} employs depth--wise separable convolutions, resulting in a compact parameter structure. It is organized into two blocks, each containing two convolutional layers, a batch normalization, dropout, and ELU activation. The MFB–CNN \cite{lee2020continuous} uses deep spatial and temporal filters, and we specifically selected this model to assess the impact of temporal features on accuracy. As shown in Table I, our proposed model achieved a superior average accuracy of 0.8801 ($\pm$0.0278), outperforming the baseline models. S6 exhibited the highest accuracy at 0.9404, while S10 showed the lowest accuracy at 0.8406. Moreover, our model demonstrated exceptional stability, with the lowest standard deviation of 0.0278, indicating robustness and consistency across all subjects.\\

\begin{figure}[t!]
\centering
\scriptsize
\centerline{\includegraphics[width=0.93\columnwidth, height=0.40\textwidth]{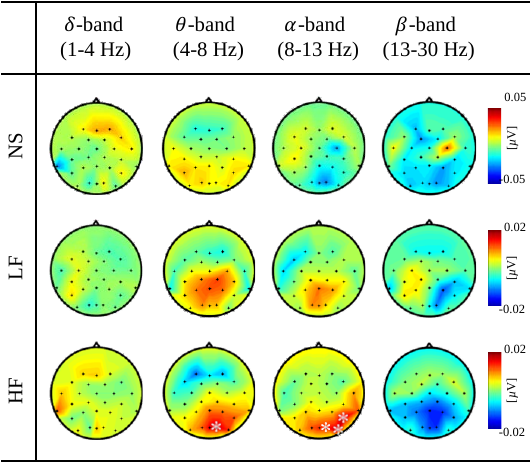}}
\caption{Scalp topographies according to the spectral bands (\textit{$\delta$}--, \textit{$\theta$}--, \textit{$\alpha$}--, and \textit{$\beta$}--bands) for a representative subject (S6). The locations of channels with the statistical significance are indicated as grey `$\ast$' (*: \textit{p}$<$0.05).}
\end{figure}

\subsection{Neurophysiological Analysis from EEG Signals}
In our previous study \cite{lee2020continuous}, we conducted a detailed analysis of various mental states by dividing the frequency spectrum into four bands (\textit{$\delta$}--, \textit{$\theta$}--, \textit{$\alpha$}--, and \textit{$\beta$}--bands) and the brain regions into three areas (temporal, central, and parietal regions). We presented scalp topographies based on the spectral bands of EEG signals for a representative subject (S6). These topographies depicted the grand--average band power, visualizing brain activation across different mental states (NS, LF, and HF). The amplitude was calculated for all EEG channels and each frequency band (\textit{$\delta$}--, \textit{$\theta$}--, \textit{$\alpha$}--, and \textit{$\beta$}--bands). Fig. 3 revealed that the amplitude varied significantly across spectral bands and brain regions. As fatigue increased, the amplitude in the \textit{$\theta$}-- and \textit{$\alpha$}--bands increased in the occipital region. Notably, the regions marked with a grey asterisk ($\ast$) denote channels where statistical significance was observed (\textit{p}\textless{}0.05). However, no distinct patterns were observed in the \textit{$\delta$}-- and \textit{$\beta$}--bands.\\

\section{CONCLUSION}
Among the numerous factors responsible for aviation accidents, pilots' mental state is a critical determinant. Given its direct impact on passenger safety, there is a pressing need for the accurate detection of pilots' mental states. The detection of different mental states through EEG signals remains the significant challenge within BCI field. Precise classification of mental states is particularly important to prevent human--induced accidents. This study introduces a novel model designed to classify fatigue levels using EEG signals. Our proposed model exhibits a superior accuracy in fatigue classification compared to conventional models. By acquiring fatigue--related EEG data from pilots, this research aims to advance technologies in autonomous flight and driving. Future work will focus on refining our model by acquiring EEG signals corresponding to various abnormal mental states using modified experimental protocols, with the goal of applying our research to real--world environments.\\

\bibliographystyle{IEEEtran}
\bibliography{REFERENCE}

\end{document}